\documentclass[twocolumn,showpacs,preprintnumbers,amsmath,amssymb]{revtex4}

\usepackage{graphicx}
\usepackage{dcolumn}
\usepackage{bm}

\begin{document}

\title{Quantum Teleportation of Superposition State for Squeezed States }

\author{Xin-Hua Cai }
\affiliation{Department of Physics, Changde Normal College,
  Changde 415000, Hunan Province, China\\and Department of Physics, Hunan Normal University,
Changsha 410081, China\ }
\author{Le-Man Kuang$^{\dagger}$}
 \affiliation{Department of Physics, Hunan Normal
University, Changsha 410081, China\ }
\begin{abstract}
This paper proposes a scheme for teleporting an arbitrary coherent
superposition state of two equal-amplitude and opposite-phase
squeezed vacuum states (SVS) via a symmetric 50/50 beam splitter
and photodetectors. It is shown that the quantum teleportation
scheme has the successful probability 1/4. Maximally entangled
SVS's are used as quantum channel for realizing the teleportation
scheme. It is shown that if an initial quantum channel is in a
pure but not maximally entangled SVS, the quantum channel may be
distilled to a maximally entangled SVS through entanglement
concentration.
\end{abstract}

\pacs{03.67.Hk, 03.67.-a, 03.65.Ud}
\maketitle

Quantum teleportation is a process by which a quantum state of a
system is transported from one location to another. The state of
the system is destroyed in one place and a perfect replica is
created at a distant site by making measurements on the system and
classical communication of the results to the remote location. The
original proposals for quantum teleportation \cite{a,b} focused on
teleporting quantum states of a system with a finite-dimensional
(discrete variable) state space, such as the two polarizations of
a photon or the discrete levels of an atoms. Discrete-variable
teleportation has been demonstrated experimentally in optical
systems \cite{c} and liquid-state nuclear magnetic resonance
systems \cite{d}.

Recently, quantum teleportation has been extended to continuous
variables corresponding to states of infinite-dimensional
(continuous-variable) systems \cite{e,f} such as optical fields or
the motion of massive particle. In particular following the
theoretical proposal of Ref. \cite{f}, continuous-variable
teleportation has been realized for coherent states of a light
field \cite{g} by using entangled two-mode squeezed optical beams
produced by parametric down-conversion in a sub-threshold optical
parametric oscillator. Although coherent states are continuous and
nonorthogonal states, they are very close to classical states. A
real challenge for quantum teleportation is to teleport truly
nonclassical states like squeezed states,  quantum superposition
states and entangled states. van Enk and Hirota \cite{h} proposed
a scheme to faithfully teleport a superposition state of two
coherent states with equal amplitudes and opposite-phases though a
quantum channel described by a maximally entangled coherent state
with one ebit of entanglement.

In this paper we propose a scheme to teleport a superposition
state of two SVS's with equal amplitudes and opposite phases using
a beam splitter and photodetectors.

In the teleportation scheme one needs certain types of entangled
states. We consider the following entangled squeezed states
defined by (ESS's)
\begin{eqnarray}
|\Phi\rangle_{\pm}=(|\xi\rangle_{1}|\xi\rangle_{2}\pm|-\xi\rangle_{1}
|-\xi\rangle_{2})/\sqrt{N_{\pm}},
\end{eqnarray}
\begin{eqnarray}
|\Psi\rangle_{\pm}=(|\xi\rangle_{1}|-\xi\rangle_{2}\pm|-\xi\rangle_{1}
|\xi\rangle_{2})/\sqrt{N_{\pm}},
\end{eqnarray}
where both $|\xi\rangle$ and $|-\xi\rangle$ are the single-mode
squeezed vacuum states
\begin{equation}
|\xi\rangle_1=\hat{S}_1(\xi)|0\rangle, \hspace{0.6cm}
|\xi\rangle_2=\hat{S}_2(\xi)|0\rangle,
\end{equation}
with the single-mode squeezing operators
\begin{eqnarray}
\hat{S}_1(\xi)&=&\exp\left(-\frac{\xi}{2}\hat{a}^{\dagger 2}_1 +
\frac{\xi^*}{2}\hat{a}^2_1\right ), \\
\hat{S}_2(\xi)&=&\exp\left(-\frac{\xi}{2}\hat{a}^{\dagger 2}_2 +
\frac{\xi^*}{2}\hat{a}^2_2\right ).
\end{eqnarray}
In Eqs. (1) and (2)  the normalization factors are given by
\begin{eqnarray}
N_{\pm}=2(1\pm k_{\xi}^{2}),
\end{eqnarray}
with
$k_{\xi}=\langle\xi|-\xi\rangle=\sqrt{\frac{sech^{2}r}{1+tanh^{2}r}}$
being the overlap of the two squeezed vacuum states $|\xi\rangle$
and $|-\xi\rangle$.

It has been shown that the ESS's in (1) and (2) can be produced
via beam splitters and squeezed-vacuum-state phase-shift
operations \cite{i}. Entanglement in these ESS's can be measured
by the partial entropy of each subsystem. In order to calculate
the partial entropy, we express the ESS's in terms of two
orthogonal
Hilbert space vectors  $|0\rangle=\left(\begin{array}{c} 1\\
0\end{array}\right)$ and $|1\rangle=\left(\begin{array}{c} 0\\
1\end{array}\right)$ as
$|\xi\rangle=|0\rangle$,$|-\xi\rangle=M|1\rangle +
k_{\xi}|0\rangle)$ with $M=\sqrt{1-k_{\xi}^{2}}$. Under these
basic vectors, the ESS's (1), (2) can be rewriten as
\begin{eqnarray}
|\Phi'\rangle_{\pm}&=&[(1\pm
k_{\xi}^{2})|0\rangle_{1}|0\rangle_{2}\pm
M^{2}|1\rangle_{1}|1\rangle_{2}\nonumber\\&&\pm M
k_{\xi}|1\rangle_{1}|0\rangle_{2}\pm
{k_{\xi}M|0\rangle_{1}|1\rangle_{2}]/\sqrt{N_{\pm}},}
\end{eqnarray}
\begin{eqnarray}
|\Psi'\rangle_{\pm}&=&[(k_{\xi}\pm
k_{\xi})|0\rangle_{1}|0\rangle_{2}+
M|0\rangle_{1}|1\rangle_{2}\nonumber\\&&\pm M
|1\rangle_{1}|0\rangle_{2}] /\sqrt{N_{\pm}}.
\end{eqnarray}
Therefore, we transfer the continuous-variable state expressions
(1) and (2) to the discrete-variable  forms (7) and (8). From (7)
and (8) it is easy to get the partial entropy of the subsystem 1
or 2
\begin{eqnarray}
E_{-}(\Phi)=E_{-}(\Psi)=1 ,
\end{eqnarray}
\begin{eqnarray}
E_{+}(\Phi)&=&E_{+}(\Psi)\nonumber\\&=&-\frac{(1+k_{\xi})^{2}}{2(1+k_{\xi}^{2})}
\log_{2}\frac{(1+k_{\xi})^{2}}{2(1+k_{\xi}^{2})}\nonumber\\
&&-\frac{(1-k_{\xi})^{2}}{2(1+k_{\xi}^{2})}
\log_{2}\frac{(1-k_{\xi})^{2}}{2(1+k_{\xi}^{2})},
\end{eqnarray}
which indicate that entangled SVS's $|\Phi\rangle_{-}$ and
$|\Psi\rangle_{-}$ are maximally entangled squeezed-vacuum states
which can be used as quantum channels for continuous-variable
quantum teleportation.

Let us briefly review the action of a beam splitter on
squeezed-vacuum states before we present our teleportation scheme.
The lossless symmetric 50/50 beam splitter is described by a
unitary transformation
\begin{eqnarray}
\hat{U}_{1,2}=exp[i\frac{\pi}{4}(\hat{a}_{1}^{\dagger}\hat{a}_{2}+\hat{a}_{2}^{\dagger}\hat{a}_{1})]
\end{eqnarray}
where $\hat{a}_{1}$ and $\hat{a}_{2}$ are the annihilation
operators for the two light beams entering the two input ports,
respectively. Let $\hat{A}_{1}$ and $\hat{A}_{2}$ denote the
annihilation operators for the two light beams leaving the two
output ports for the 50/50 beam splitter, the photon annihilation
and creation operators of the output modes $\hat{A}_{i}$ and
$\hat{A}_{i}^{+}$, respectively, can then be obtained using the
well-known input-output relations
\begin{eqnarray}
\hat{A}_{1}=\frac{1}{\sqrt{2}}(\hat{a}_{1}+i\hat{a}_{2}),
\hat{A}_{2}=\frac{1}{\sqrt{2}}(\hat{a}_{2}+i\hat{a}_{1}).
\end{eqnarray}

Assume that before interacting with the beam splitter the two
input light beams are in squeezed vacuum states
\begin{equation}
|\Psi_{in}\rangle=\hat{S}_1(\xi)\hat{S}_2(\xi)|0,0\rangle,
\end{equation}
where $\hat{S}_1(\xi)$ and $\hat{S}_2(\xi)$ are given by Eqs. (4)
and (5), respectively.  After interacting with the beam splitter,
the state vector of the output light beams becomes
\begin{eqnarray}
|\Psi_{out}\rangle&=&\exp[-\frac{1}{4}(\xi_{1}-\xi_{2})\hat{A}_{1}^{\dagger2}
+\frac{1}{4}(\xi_{1}^{*}-\xi_{2}^{*})\hat{A}_{1}^{2}\nonumber\\
&+&\frac{1}{4}(\xi_{1}-\xi_{2})\hat{A}_{2}^{\dagger2}
-\frac{1}{4}(\xi_{1}^{*}-\xi_{2}^{*})\hat{A}_{2}^{2}\nonumber\\
&-&\frac{i}{2}(\xi_{1}+\xi_{2})\hat{A}_{1}^{\dagger}\hat{A}_{2}^{\dagger}
-\frac{i}{2}(\xi_{1}^{*}+\xi_{2}^{*})\hat{A}_{1}\hat{A}_{2}]|0,0\rangle.\nonumber\\
\end{eqnarray}
For later use, let us set $\xi_{i}=r_{i}e^{i\varphi_{i}}$ with
$r_{i}$ being the squeezing amplitude and $\varphi_{i}$ the
squeezing angle, respectively, and  consider the following two
cases which play a key role in our teleportation scheme.

 Case 1. The two input light beams have the same squeezing
 amplitudes and phases, i.e;
 $r_{1}=r_{2}=r$, and $\varphi_{1}=\varphi_{2}$. In this case, from
 Eq.(14) it is easy to know that the state of the output light
 beams is simply a two-mode squeezed state
\begin{eqnarray}
|\Psi_{out}\rangle=\exp[-re^{i(\varphi+\frac{\pi}{2})}
\hat{A}_{1}^{\dagger}\hat{A}_{2}^{\dagger}+re^{-i(\varphi+\frac{\pi}{2})}
\hat{A}_{1}\hat{A}_{2}]|0,0\rangle.\nonumber\\
\end{eqnarray}

Case 2. The two input light beams have the same squeezing
amplitudes but a phase difference $\pi$, i.e., $r_{1}=r_{2}=r$ and
 $\varphi_{2}-\varphi_{1}=\pi$. In this case, the output state of
the beam splitter becomes a direct product of two single-mode
squeezed vacuum states
\begin{eqnarray}
|\Psi_{out}\rangle&=&\exp(-\frac{1}{2}re^{i\varphi}
\hat{A}_{1}^{\dagger2}+\frac{1}{2}re^{-i\varphi}\hat{A}_{1}^{2})\nonumber\\
&\times&\exp(-\frac{1}{2}re^{i(\varphi+\pi)}\hat{A}_{2}^{\dagger2}
+\frac{1}{2}re^{-i(\varphi+\pi)}\hat{A}_{2}^{2})|0,0\rangle.\nonumber\\
\end{eqnarray}

As is known, it is possible to generate a superposition state of
the two SVS's $|\xi\rangle$ and $|-\xi\rangle$ from a squeezed
vacuum states $|\xi\rangle$ propagating through a nonlinear medium
\cite{j}. Now let us assume that by using the quantum channel
described by the maximally entangled state $|\Phi\rangle_{-}$,
Alice wants to teleport Bob a coherent superposition state of the
two squeezed vacuum states
\begin{eqnarray}
|\Psi\rangle_{0}=(C_{+}|\xi\rangle_{0}+C_{-}|-\xi\rangle_{0})/\sqrt{N_{\Psi}},
\end{eqnarray}
where $C_{+}$ and $C_{-}$ are any complex numbers, $N_{\Psi}$
normalization factor given by
\begin{eqnarray}
N_{\Psi}=|C_{+}|^{2}+|C_{-}|^{2}+2k_{\xi}Re(C_{+}C_{-}^{*}).
\end{eqnarray}
Then the initial state of the whole system consisting of system 0,
1,  and 2 is given by
\begin{eqnarray}
|\Phi\rangle_{012}&=&|\Psi\rangle_{0}|\Phi\rangle_{-}\nonumber\\
&=&\frac{1}{\sqrt{N_{\Psi}N_{-}}}(C_{+}|\xi\rangle_{0}|\xi\rangle_{1}|\xi\rangle_{2}
-C_{+}|\xi\rangle_{0}|-\xi\rangle_{1}|-\xi\rangle_{2}\nonumber\\
&+&C_{-}|-\xi\rangle_{0}|\xi\rangle_{1}|\xi\rangle_{2}
-C_{-}|-\xi\rangle_{0}|-\xi\rangle_{1}|-\xi\rangle_{2}).\nonumber\\
\end{eqnarray}
Note that at this time the system 0 and 1 at the sender's side and
system 2 at the receiver's side. We apply the beam splitter
transformation $\hat{U}_{01}$ to the initial state (19), the state
after the transformation becomes
\begin{eqnarray}
|\Phi'\rangle_{012}&=&\frac{1}{\sqrt{N_{\Psi}N_{-}}}
[C_{+}\hat{S}_{0,1}(re^{i(\varphi+\frac{\pi}{2})})
|0,0\rangle_{01}\otimes|\xi\rangle_{2}\nonumber\\
&&-C_{+}\hat{S}_{0}(re^{i\varphi})\hat{S}_{1}(-re^{i\varphi})|0,0\rangle_{01}
\otimes|-\xi\rangle_{2}\nonumber\\
&&+C_{-}\hat{S}_{0}(-re^{i\varphi})\hat{S}_{1}(re^{i\varphi})|0,0\rangle_{01}
\otimes|\xi\rangle_{2}\nonumber\\
&&-C_{-}\hat{S}_{0,1}(re^{i(\varphi-\frac{\pi}{2})})
|0,0\rangle_{01}\otimes|-\xi\rangle_{2}],
\end{eqnarray}
where the two-mode squeezed operator is given by
\begin{eqnarray}
\hat{S}_{0,1}(\xi)=\exp(-\xi\hat{A}_{0}^{+}\hat{A}_{1}^{+}+\xi^{*}\hat{A}_{1}\hat{A}_{0}).
\end{eqnarray}
 Subsequently, Alice performs photon number measurements on system 0 and 1 on her
side. From Eq.(20) it can be seen that for system 0 and 1 the
second and third term on the RHS are single-mode SVS's which
contain only even-number photon states in their number-state
expansions, while the first and fourth terms are two-mode squeezed
states in which each mode has the same photon numbers with both
even and odd photons in the photon number-state representation.

After Alice measures an odd number of photons in either mode of
systems 0 and  1, the state of the whole system collapses into the
following state
\begin{eqnarray}
|\Phi''\rangle_{012}&=&\frac{1}{\sqrt{N_{\Psi}N_{-}}}
sech r[e^{i(\varphi-\frac{\pi}{2})}\tanh r]^{2n+1}\nonumber\\
&\times&|2n+1,2n+1\rangle_{01}(C_{+}|\xi\rangle_{2}+C_{-}|-\xi\rangle_{2})\nonumber\\
\end{eqnarray}
which indicates that Alice transmits the coherent superposition of
the two squeezed states $|\Psi\rangle_{0}$ given in Eq.(17) to
Bob. The probability of finding an odd number of photons from the
output state of the beam splitter (20) is given by
\begin{eqnarray}
P(2n+1)=\frac{1}{N_{-}}sech^{2}r(\tanh r)^{2(2n+1)},
\end{eqnarray}
which is independent of the state to be teleported, hence we
obtain the perfect teleportation for the odd number of photons.
Then the successful probability of the teleportation scheme is
given by
\begin{eqnarray}
P=\sum_{n=0}^{\infty} P(2n+1).
\end{eqnarray}
These summation can be performed easily and the result is found to
be independent of the squeezing amplitude $r$,
\begin{eqnarray}
P=\frac{1}{4}.
\end{eqnarray}

In above teleportation scheme, a highly entangled state for the
quantum channel plays a key role. If the initially prepared
quantum channel is in a pure but not maximally ESS, the quantum
channel may be distilled to a maximally entangled squeezed state
before using it for quantum teleportation through entanglement
concentration\cite{k,l}. Here we show that for an ESS the
entanglement concentration may be simply realized by using a beam
splitter and photodetectors.

Suppose an ensemble of a partially entangled pure squeezed state
\begin{eqnarray}
|\Phi\rangle_{12}=\frac{1}{\sqrt{N_{\eta}}}(\cos\eta|\xi\rangle_{1}|-\xi\rangle
_{2}-\sin\eta|-\xi\rangle_{1}|\xi\rangle_{2}),
\end{eqnarray}
from which we want to distill a subensemble of a maximally
entangled squeezed state. In Eq.(26)  $N_{\eta}$ is a normalized
factor given by
\begin{eqnarray}
N_{\eta}=1-\sin(2\eta)\frac{sech^{2}r}{1+\tanh^{2}r},
\end{eqnarray}
where the phase factor $\eta$, $0<\eta<\pi/2$, determines the
degree of entanglement for the state (26). It is obvious that the
partially entangled pure squeezed state $|\Phi\rangle_{12}$
includes two ESS's $|\Phi\rangle_{-}$ and $|\Psi\rangle_{-}$ in
Eq.(1) and (2) as two particular cases.

After sharing a quantum channel between Alice and Bob
$|\Phi\rangle_{12}$ Alice prepares a pair of particles which are
of the same entangled squeezed state $|\Phi\rangle_{34}$. The
initial state of the whole system consisting of the four
subsystems is then given by
\begin{eqnarray}
|\Psi\rangle_{1234}=|\Phi\rangle_{12}\otimes|\Phi\rangle_{34}.
\end{eqnarray}
For convenience, now we assume mode 1 at the Bob's side and modes
2,3,4 at the Alice's side, then state (28) can be explicitly
written as
\begin{eqnarray}
|\Psi\rangle_{1234}&=&\frac{1}{N_{\eta}}[\cos^{2}\eta|\xi\rangle_{1}
\otimes|-\xi\rangle_{2} \otimes|\xi\rangle_{3}
\otimes|-\xi\rangle_{4}\nonumber\\&-&\frac{1}{2}\sin(2\eta)|\xi\rangle_{1}
\otimes|-\xi\rangle_{2} \otimes|-\xi\rangle_{3}
\otimes|\xi\rangle_{4}\nonumber\\&-&\frac{1}{2}\sin(2\eta)|-\xi\rangle_{1}
\otimes|\xi\rangle_{2} \otimes|\xi\rangle_{3}
\otimes|-\xi\rangle_{4}\nonumber\\&+&\sin^{2}\eta|-\xi\rangle_{1}
\otimes|\xi\rangle_{2} \otimes|-\xi\rangle_{3}
\otimes|\xi\rangle_{4}].\nonumber\\
\end{eqnarray}
Subsequently, Alice let modes 2 and modes 3 enter the input ports
of a beam splitter. After interacting with the beam splitter, the
state of the whole system becomes
\begin{eqnarray}
|\Psi'\rangle_{1234}&=&\frac{1}{N_{\eta}}[\cos^{2}\eta|\xi\rangle_{1}
\otimes|-\xi\rangle_{2}\otimes|\xi\rangle_{3}\otimes|-\xi\rangle_{4}
\nonumber\\&-&\frac{1}{2}\sin(2\eta)|\xi\rangle_{1}\otimes\hat{S}_{23}
(re^{i(\varphi-\frac{\pi}{2})})
|0,0\rangle_{23}\nonumber\\&&\otimes|\xi\rangle_{4}
\nonumber\\&-&\frac{1}{2}\sin(2\eta)|-\xi\rangle_{1}\otimes\hat{S}_{23}
(re^{i(\varphi+\frac{\pi}{2})})|0,0\rangle_{23}\nonumber\\&&\otimes
|-\xi\rangle_{4}\nonumber\\&+&\sin^{2}\eta
|-\xi\rangle_{1}\otimes|\xi\rangle_{2}
\otimes|-\xi\rangle_{3}\otimes|\xi\rangle_{4}],\nonumber\\
\end{eqnarray}
where the two-mode squeezed operator is given by (21).

Now Alice performs photon number measurements for mode 2 and 3,
when the results of measurements are odd number of photon, from
Eq.(30) it can be seen that modes 1 and 4 collapse to a maximally
ESS
\begin{eqnarray}
|\Psi\rangle_{14}=\frac{1}{\sqrt{N_{-}}}(|\xi\rangle_{1}\otimes
|\xi\rangle_{4}-|-\xi\rangle_{1}\otimes|-\xi\rangle_{4}).
\end{eqnarray}
From Eq.(30), we can obtain the probability of finding $2n+1$
photons in modes 2 and 3 to be
\begin{eqnarray}
P(2n+1)=\frac{N_{-}}{4N_{\eta}^{2}}\sin^{2}(2\eta)sech^{2}r(\tanh
r)^{2(2n+1)},
\end{eqnarray}
which leads the probability of obtaining the maximally ESS (31) to
be
\begin{eqnarray}
P&=&\sum_{n=0}^{\infty}P(2n+1)\nonumber\\&=&\frac{1}{4}\sin^{2}(2\eta)
\frac{N_{-}\tanh^{2}r}{N_{\eta}^{2}(1+\tanh^{2}r)},
\end{eqnarray}
which implies that no matter how small the initial entanglement
is, it is possible to distill some maximally entangled coherent
channels from partially entangled pure channels.

In summary, we have present a scheme to optically teleport an
arbitrary coherent superposition state of two equal-amplitude and
opposite-phase SVS's via a symmetric 50/50 beam splitter and
photodetectors with the successful probability 1/4. In particular,
a perfect teleportation of a single-mode SVS can be obtained when
one of the two coefficients in the coherent superposition state of
two SVS's vanishes.  In our scheme, maximally entangled SVS's are
used as quantum channel for realizing the quantum teleportation
scheme. We have also shown how  to distill a maximally entangled
SVS from an initial quantum channel in a pure but not maximally
entangled SVS through entanglement concentration.

\acknowledgments

This work was supported in part the National "973" Research Plan
of China, the Natinal Natural Science Foundation of China, and the State
Education Ministry of China.

$^\dagger$Corresponding author.


\begin{references}
\bibitem {a} C. H. Bennett, G. Brassard, C. Cr$\acute{e}$peau, R. Jozsa,
A. Peres, and W. K. Wootters, Phys. Rev. Lett. {\bf 70}, 1895
(1993).
\bibitem {b} L. Davidovich, N. Zagury, M. Brune, J. M. Raimond,
and S. Harroche, Phys. Rev. A {\bf 50}, R895 (1994).
\bibitem {c} D. Bouwmeester, J. W. Pan, K. Mattle, M. Eibl, H.
Weinfurter, and A. Zeilinger, Nature {\bf 390} 575 (1997).
\bibitem {d} M. A. Nielsen, E. Knill, and R. Laflamme, Nature {\bf 396} 52
(1998).
\bibitem {e} L. Vaidman, Phys. Rev. A {\bf 49}, 1473 (1994).
\bibitem {f} S. L. Braunstein and H. J. Kimble, Phys. Rev. Lett.
{\bf 80}, 869 (1998).
\bibitem {g} A. Furusawa, J. L. Srensen, S. L. Braunstein, C.
A.Fuchs, H. J. Kimble, and E. S. Polzik, Science {\bf 282} 706
(1998).
\bibitem {h} S. J. van Enk and O. Hirota, Phys. Rev. A {\bf 64}, 022313 (2001).
\bibitem {i} L. Zhou and L. M. Kuang, in preparation.
\bibitem {j} B. C. Sanders, Phys. Rev. A {\bf 39}, 4284 (1989).
\bibitem {k} C. H. Bennett, S. Popescu, B. Schumacher, J. A.
Smolin, and W. K. Wootters, Phys. Rev. Lett. {\bf 76}, 722 (1996).
\bibitem {l} C. H. Bennett, H. J. Bernstein, S. Popescu, and B.
Schumacher, Phys. Rev. A {\bf 53}, 2046 (1996).
\end{references}
\end{document}